\documentclass[twocolumn,prb,aps,superscriptaddress,showpacs,amsmath,amssymb]{revtex4}
\usepackage{amsfonts}
\usepackage{bbm}

\usepackage{graphicx}% Include figure files
\usepackage{dcolumn}% Align table columns on decimal point
\usepackage{bm}% bold math
\begin{document}

\title{Delocalization and scaling properties of low-dimensional quasiperiodic systems}
\author{Ai-Min Guo}
\email{aimin.guo218@gmail.com}
\affiliation{Institute of Physics, Chinese Academy of Sciences, Beijing 100190, China}
\author{X. C. Xie}
\affiliation{International Center for Quantum Materials, Peking
University, Beijing 100871, China} \affiliation{Collaborative
Innovation Center of Quantum Matter, Beijing 100871, China}
\author{Qing-feng Sun}
\email{sunqf@pku.edu.cn} \affiliation{International Center for
Quantum Materials, Peking University, Beijing 100871, China}
\affiliation{Collaborative Innovation Center of Quantum Matter,
Beijing 100871, China}

\begin{abstract}

In this paper, we explore the localization transition and the scaling properties of both quasi-one-dimensional and two-dimensional quasiperiodic systems, which are constituted from coupling several Aubry-Andr\'{e} (AA) chains along the transverse direction, in the presence of next-nearest-neighbor (NNN) hopping. The localization length, two-terminal conductance, and participation ratio are calculated within the tight-binding Hamiltonian. Our results reveal that a metal-insulator transition could be driven in these systems not only by changing the NNN hopping integral but also by the dimensionality effects. These results are general and hold by coupling distinct AA chains with various model parameters. Furthermore, we show from finite-size scaling that the transport properties of the two-dimensional quasiperiodic system can be described by a single parameter and the scaling function can reach the value 1, contrary to the scaling theory of localization of disordered systems. The underlying physical mechanism is discussed.

\end{abstract}


\maketitle

\section{Introduction\label{sec1}}

Since the original prediction by Anderson that the disorder could lead to the absence of electron diffusion in imperfect crystals,\cite{apw1} Anderson localization has always been one of the most fascinating phenomena in condensed-matter physics and much progress has been achieved in this research field.\cite{ae1,ef1,ae2,js1} In fact, Anderson localization is a ubiquitous quantum phenomenon and has been directly observed in a variety of disordered systems.\cite{st1,ly1,hh1,bj,rg1,fs1,ly2,kss1,jf1}

On the other hand, the complementary subject to explore metallic states in low-dimensional disordered systems arises and has attracted extensive attention among the physics community, in the context of electron transport and direct application of realistic materials which are always disordered. For instance, the extended states have been reported in one-dimensional (1D) disordered systems when the short- or long-range correlations are incorporated.\cite{dm1,bv1,ifm1,ku1,ifm2} Another celebrated example is the Aubry-Andr\'{e} (AA) model\cite{aa1}
\begin{eqnarray}
[E- W \cos(2\pi \alpha n)] \varphi_n= t (\varphi_{n-1} + \varphi _{n+1}), \label{eq1}
\end{eqnarray}
where $E$ is the Fermi energy, $W$ is the strength of the on-site potential, $\alpha$ is an irrational number and is incommensurate with the lattice, $t$ is the nearest-neighbor hopping integral, and $\varphi_n$ is the amplitude of the wavefunction at the $n$th site. Because $\alpha$ is irrational, the on-site potential displays quasiperiodicity and the AA model can be regarded as a quasiperiodic (QP) system.

Let us consider the commensurate case of $\alpha=p/q$ with $p$ and $q$ being co-prime integers. It is obvious that the resultant energy spectrum of Eq.~(\ref{eq1}) consists of $q$ subbands and all electronic states are described by Bloch's theorem. Since an irrational number can be viewed as the ratio of two infinite integers, the energy spectrum of the AA model splits into infinite subbands, leading to a highly fractal band structure. Besides, it has been claimed that all electronic states of the AA model are delocalized when $W < 2t$ and are exponentially localized with Lyapunov exponent $\ln (W/2t)$ when $W>2t$.\cite{aa1} This metal-insulator transition (MIT) in parameter space has been verified experimentally in both Bose-Einstein condensates and photonic lattices. Roati {\it et al.} have demonstrated this phase transition for noninteracting ultracold atoms in optical lattices by studying the transport properties and both spatial and momentum distributions.\cite{rg1} Lahini {\it et al.} have directly measured the spreading of initially narrow wave packets in photonic lattices and observed a localization phase transition at $W=2t$.\cite{ly2} These recover the interest of the electronic properties of the AA model. Sil {\it et al.} have shown that a ladder network, consisting of two identical AA chains, exhibits an MIT at multiple Fermi energies in the presence of next-nearest-neighbor (NNN) hopping.\cite{ss1} Biddle {\it et al.} have studied the localization properties of the AA chain by considering non-nearest-neighbor hopping and found the mobility edges.\cite{bj1,bj2}

\begin{figure}
\includegraphics[width=0.4\textwidth]{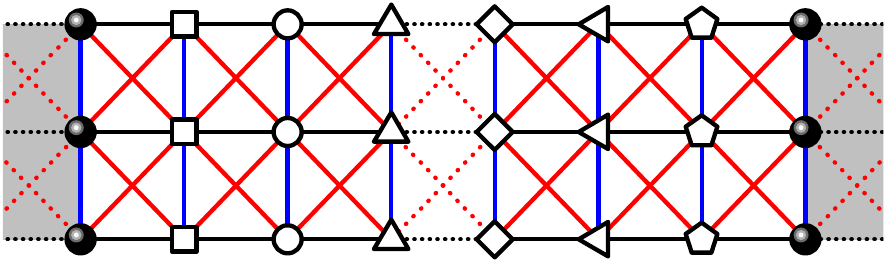}
\caption{\label{fig:one}(Color online) Schematic view of the QP system (white region with open symbols) connected by two leads (gray region with filled circles). The black, blue, and red lines denote the intrachain, interchain, and NNN hopping integrals, respectively. Here, the width of the system is $S=3$.}
\end{figure}

Since the realistic materials usually have finite width, there is a growing interest in the localization properties of quasi-1D system\cite{dm2,zw1,gam1,xhy1,kjj1,sta1,ra1} and of two-dimensional (2D) system in recent years.\cite{hm1,pa1,om1,bjh1,nk1,zyy1,xz1} In this perspective, it would be significant to investigate the localization transition and the scaling properties of both quasi-1D and 2D QP systems, which are made up from coupling several AA chains along the transverse direction, and there has already been one work along this research field.\cite{ss1} One wonders whether there still exists an MIT in these systems in parameter space and whether the scaling theory of localization remains valid in the 2D QP system. In this paper, we will answer the above two questions and Fig.~\ref{fig:one} plots the schematic illustration of the system contacted by two leads. By combining the Landauer-B\"{u}ttiker formula with recursive Green's function technique, the conductance of the QP system is calculated in the presence of NNN hopping. Our results reveal that an MIT could occur in the quasi-1D QP system by tuning not only the NNN hopping integral but also the number of the AA chains. All these results are general and still hold by coupling distinct AA chains with different strengths of the on-site potentials and other model parameters. In addition, we show from finite-size scaling that the transport properties of the 2D QP system can be described by a single parameter and the scaling function can reach the value 1, contrary to the scaling theory of localization. Finally, the MIT phase diagram is obtained and the statistical properties of the conductance are performed for the 2D QP system.

The rest of the paper is constructed as follows. In Sec.~\ref{sec2}, the theoretical model and the method are presented. In Sec.~\ref{sec3}, the conductance and the localization length are shown for different situations, and the scaling and statistical properties are also studied for the 2D QP system. Finally, the results are summarized in Sec.~\ref{sec4}.

\section{Model and Method\label{sec2}}

Within the framework of the tight-binding approximation, the Hamiltonian of the QP system connected by two leads can be written as
\begin{eqnarray}
{\cal H} = && \sum_{j,n}[\varepsilon_{j,n} c_{j,n}^\dagger c_{j,n} + t c_{j,n}^\dagger c_{j,n+1} + \lambda c_{j,n}^\dagger c_{j+1,n} \nonumber \\&& + t_d ( c_{j,n}^\dagger c_{j-1,n+1}+c_{j,n}^\dagger c_{j+1,n+1} ) + {\rm H.c.}], \label{eq2}
\end{eqnarray}
where $c_{j,n}^\dagger$ ($c_{j,n}$) creates (annihilates) an electron at lattice site $(j,n)$ of the left lead ($n<1$), of the central region ($1\leq n \leq L$, white region in Fig.~\ref{fig:one}), and of the right lead ($n>L$). Here, $j\in[1,S]$ is the chain index, $S$ is the width, and $L$ is the length of the central region. The on-site potential $\varepsilon_{j,n}$ at the central region is set as
\begin{eqnarray}
\varepsilon _{j,n}= W_j \cos (2\pi \alpha n), \label{eq3}
\end{eqnarray}
with $W_j$ the strength of the on-site potential of the $j$th chain and $\alpha$ the irrational number. $t$ is the intrachain hopping integral and is chosen as the energy unit. $\lambda$ and $t_d$ are, respectively, the interchain and NNN hopping integrals between two neighboring chains (see Fig.~\ref{fig:one}).

The Schr\"{o}dinger equation ${\cal H}| {\mathbf \Psi} \rangle= E|{\mathbf \Psi} \rangle$ in the site representation can be expressed as
\begin{eqnarray}
\begin{aligned}
(E{\mathbf I}- {\cal H}_n) {\mathbf \Psi}_n={\cal M} ({\mathbf \Psi}_{n-1} + {\mathbf \Psi}_{n+1}). \label{eq4}
\end{aligned}
\end{eqnarray}
Here, ${\mathbf I}$ is the $S\times S$ identity matrix, ${\cal H}_n$ is the sub-Hamiltonian matrix of the $n$th layer
\begin{eqnarray}
\begin{aligned}
{\cal H}_n=\left(\begin{array}{ccccc}
\varepsilon_{1,n} & \lambda & 0 & \cdots & 0 \\
\lambda & \varepsilon_{2,n} & \lambda & \ddots & \vdots \\
0 & \lambda & \varepsilon_{3,n} & \ddots & 0 \\
\vdots & \ddots & \ddots & \ddots & \lambda \\
0 & \cdots & 0 & \lambda & \varepsilon_{S,n} \\
\end{array}\right), \label{eq5}
\end{aligned}
\end{eqnarray}
${\mathbf \Psi}_n= (\psi_{1,n},\psi_ {2,n},\cdots, \psi_{S,n})^T$ with $\psi_{j,n}$ the amplitude of the wavefunction at lattice site $(j,n)$ and $T$ the transpose, and ${\cal M}$ is the sub-Hamiltonian matrix connecting two successive layers
\begin{eqnarray}
\begin{aligned}
{\cal M}=\left(\begin{array}{ccccc}
t & t_d & 0 & \cdots & 0 \\
t_d & t & t_d & \ddots & \vdots \\
0 & t_d & t & \ddots & 0 \\
\vdots & \ddots & \ddots & \ddots & t_d \\
0 & \cdots & 0 & t_d & t \\
\end{array}\right). \label{eq6}
\end{aligned}
\end{eqnarray}

At zero temperature, the Landauer conductance $G$ can be obtained from the Landauer-B\"{u}ttiker formula:\cite{ds1}
\begin{eqnarray}
G= G_0 {\rm Tr} [{\bf \Gamma}_R {\bf G}^r {\bf \Gamma}_L {\bf G}^a] \label{eq7},
\end{eqnarray}
with $G_0 ={2e^2} /h$ the quantum conductance. ${\bf \Gamma}_{ L/R} =i({\bf \Sigma}_ {L/R}^r - {\bf \Sigma}_ {L/R}^a )$ is the linewidth function, ${\bf G}^r=({\bf G}^a)^\dagger =(E {\mathbf I} - {\bf H} -{\bf \Sigma}_L^r- {\bf \Sigma}_R^r)^{-1}$ is the Green's function,  ${\bf \Sigma}_ {L/R}^r$ is the retarded self-energy due to the coupling to the left/right lead with the on-site potential being $0$,\cite{ldh1} and ${\bf H}$ is the Hamiltonian of the central region. In the numerical calculation, the conductance can be evaluated by using recursive Green's function technique for very large system size. Besides, the Lyapunov exponents $\gamma_j$'s are also calculated by the transfer-matrix method.\cite{gam1} For the $S \times L$ system, there are $S$ propagating channels along the longitudinal direction and each $\gamma_ j$ reflects the exponential decay of the corresponding channel. The smallest $\gamma_j$ is the most physically significant quantity and its inverse is the localization length $\xi$. We emphasize that the propagating channel is distinct from the chain in Eq.~(\ref{eq2}) because of the interchain hopping integral $\lambda$ and the NNN one $t_d$, and these two concepts are equivalent to each other when $\lambda=t_d=0$. The fixed boundary condition is imposed in the transverse direction. In this situation, the system is closer to the realistic materials, of which their edges usually do not connect to each other; and corrections to scaling disappear faster with increasing system size.\cite{sk1,sk2}

In the following, the irrational number is taken as $\alpha = 1 / ({10\pi})$ and the results are usually averaged over $10^4$ QP samples to reach convergence, except for specific indication in the figure caption. Each QP sample is uniformly selected from the infinite system [see Eq.~(\ref{eq3}) where the integer $n$ can be $+\infty$ theoretically] by using the sliding window strategy.\cite{gam2} For instance, the $m$th sample can be expressed as $ \varepsilon _{j,n} = W_j \cos [2\pi \alpha (n+ i*m-1)]= W_j \cos [2\pi \alpha n + \varphi_{i,m}]$ with $i$ and $m$ being positive integers. Here, the integer $i$ is chosen as $i=1$ and the results are independent on $i$ when the number of samples is sufficiently large.

\section{Results and Discussions\label{sec3}}

\subsection{Delocalization induced by the NNN hopping}

We first consider the case of all $W_j=W$ and prove analytically the NNN hopping-induced MIT in the QP system which is composed of several identical AA chains. In this regard, all ${\cal H}_n$'s and ${\cal M}$ can be diagonalized by applying a single unitary matrix $\mathbf U$ through ${\cal P}_n=\mathbf U^{\dagger} {\cal H}_n \mathbf U$ and ${\cal Q} = \mathbf U^{\dagger} {\cal M} \mathbf U $.\cite{gam3} Then, Eq.~(\ref{eq4}) can be decoupled into following Schr\"{o}dinger equations:
\begin{eqnarray}
(E- {\mathbf \nu}_{k,n}) {\mathbf \phi}_{k,n}=\mu_k ({\mathbf \phi}_{k,n-1} + {\mathbf \phi}_{k,n+1}), \label{eq8}
\end{eqnarray}
with $k\in [1,S]$, $(\phi_{1,n},\phi_{2,n},\cdots, \phi_{S,n})^T =\mathbf U^{\dagger} {\mathbf \Psi}_n$, and $\nu_{k,n}$ and $\mu_k$ the $k$th eigenvalue of ${\cal H}_n$ and $\cal M$, respectively.
\begin{eqnarray}
\nu_{k,n}= W\cos(2\pi \alpha n)+ 2\lambda \cos \frac {k\pi} {S+1}  \label{eq9}
\end{eqnarray}
and
\begin{eqnarray}
\mu_k= t+ 2t_d \cos \frac {k\pi} {S+1}.  \label{eq10}
\end{eqnarray}
Accordingly, the QP system can be transformed into $S$ isolated AA chains with identical strength $W$ of the on-site potential [Eq.~(\ref{eq9})] and different hopping integrals [Eq.~(\ref{eq10})]. These decoupled chains are exactly the propagating channels as mentioned above. One can see that the interchain hopping integral only determines the positions of the energy bands of the decoupled chains and the energy bands will be farther away even separated from each other for greater $\lambda$. In contrast, the NNN hopping integral only contributes to the hopping integrals of the decoupled chains, and some of the decoupled chains will be more delocalized with increasing $t_d$ and contrarily the others become more localized. Besides, we can also demonstrate the mobility edges from Eqs.~(\ref{eq9}) and~(\ref{eq10}) for the specific case $t_d=t$.\cite{ss1} It is clear that when anyone of the decoupled chains has extended states, the QP system shows metallic behavior. Since the most conducting decoupled chain is the first one with $k=1$, the critical condition to observe the MIT in the QP system is\cite{aa1}
\begin{eqnarray}
W_c=2( t+ 2 t_d \cos \frac {\pi} {S+1}).  \label{eq11}
\end{eqnarray}
When $W<W_c$ there will be at least one delocalized decoupled chain and the system exhibits metallic behavior, whereas $W>W_c$ all of the decoupled chains are localized and the system is insulating. Although the analytical argument is performed on the special case of all $W_j=W$, we believe that the NNN hopping-induced MIT is general for other quasi-1D QP systems by coupling distinct AA chains.

\begin{figure}
\includegraphics[width=0.45\textwidth]{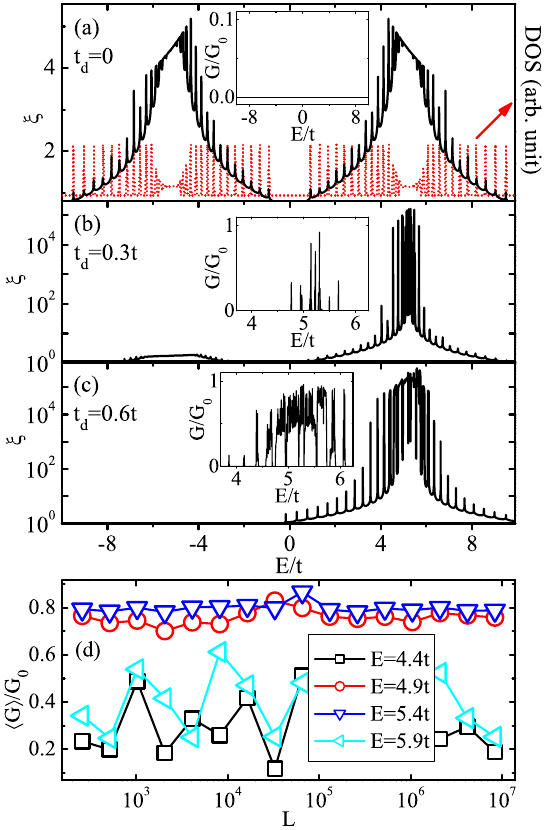}
\caption{\label{fig:two}(Color online) Energy-dependent localization length $\xi$ of the two-leg ladder for (a) $t_d=0$ (solid line), (b) $t_d=0.3t$, and (c) $t_d=0.6t$ with $L=10^5$. The insets show the corresponding conductance $G$ and the dotted line in (a) denotes the DOS. The above results are performed for a single QP sample. (d) Length-dependent averaged conductance $\langle G \rangle$ for typical values of $E$ with $t_d=0.6t$, which is obtained from $10^4$ QP samples. Other parameters are $S=2$, $\lambda= 5t$, $W_1=t$, and $W_2 =4t$.}
\end{figure}

To explore the generality of the NNN hopping-induced MIT in the QP system, we study the localization properties of the two-leg ladder model as an example, by coupling two distinct AA chains with different $W_j$'s. In this situation, it is difficult to provide analytical results and the numerical calculations are performed instead. Figures~\ref{fig:two}(a), \ref{fig:two}(b), and \ref{fig:two}(c) show the localization length $\xi$ for $t_d=0$, $0.3t$, and $0.6t$, respectively, as a function of energy $E$ with $W_1=t$ and $W_2=4t$, while the insets display the corresponding Landauer conductance $G$. Here, $\xi$ is averaged within a very small energy interval of $10^{-4}$ to avoid numerical fluctuations. Although the localization properties of the two-leg ladder strongly depend upon the NNN hopping integral $t_d$ (discussed later), one can see several general features that are irrespective of $t_d$. Firstly, all the energy spectra consist of two isolated subbands, because of the repulsion effects between the two chains driven by the large interchain hopping integral $\lambda=5t$ [Eq.~(\ref{eq9})]. And these two subbands are exactly the propagating channels of the two-leg ladder. Secondly, there exist many sharp peaks in the curve $\xi$-$E$ due to the highly fractal energy band of the AA model. This feature can be further identified in the density of states (DOS) [see the dotted line in Fig.~\ref{fig:two}(a)] which is obtained by diagonalizing the Hamiltonian of the central region, and the quantized energy levels coincide with the peaks in the curve $\xi$-$E$ [Fig.~\ref{fig:two}(a)]. Here, we emphasize that the DOS profile does not change with the system length $L$ when the two-leg ladder is sufficiently long. The origin of the extremely fractal energy spectrum can be also understood as follows: the AA model can be projected from the 2D square lattice which develops degenerate Landau energy levels in the presence of a perpendicular magnetic field.\cite{hdr1}

In the absence of $t_d$, the two subbands are almost symmetric with respect to $E=0$ and the two-leg ladder exhibits insulating behavior with $\xi \ll L$ and $G=0$ [see the solid line and the inset of Fig.~\ref{fig:two}(a)], although all electronic states of the first AA chain are delocalized ($W_1=t$), consistent with previous study that the system composed of one disordered chain and another completely ordered chain is localized.\cite{zw1} This is attributed to the fact that when an electron is injected into the first chain, it will be continuously scattered from larger (localized) potentials of the second chain, owing to the interchain hopping integral. The longer the system is, the more the scattering the electron suffers. Accordingly, $G$ will be exponentially declined to zero with increasing $L$, a typical feature of Anderson localization.

While in the presence of $t_d$ that could lead to electron-hole asymmetry, the two subbands become asymmetric and opposite behaviors could be observed with increasing $t_d$. For the left subband, its bandwidth decreases monotonically from $8.6t$ at $t_d=0$ to $6.4t$ at $t_d=0.6t$ and $\xi$ is gradually declined by increasing $t_d$, and hence the electronic states become more localized. In contrast, the bandwidth of the right subband is enhanced from $8.6t$ at $t_d=0$ to $10.8t$ at $t_d=0.6t$ and $\xi$ increases with $t_d$, and nonzero conductance could emerge in the region of large $t_d$ [see the insets of Figs.~\ref{fig:two}(b) and \ref{fig:two}(c)]. This phenomenon can be explained as follows. In fact, the NNN hopping integral could produce additional two propagating channels with opposite sign of hopping integrals [Eq.~(\ref{eq10})] and thus affects the transport properties of the two-leg ladder by mixing these channels with the original ones at $t_d=0$. When the hopping integrals of the additional and original propagating channels are of different signs, the effective hopping integral will be reduced and leads to the shrink of both the bandwidth and the localization length. Contrarily, when they have identical sign, the effective hopping integral can be enhanced and both the bandwidth and the localization length can be increased, leading to a possible MIT in the two-leg ladder. Actually, the electronic states of finite conductance observed in the insets of Figs.~\ref{fig:two}(b) and \ref{fig:two}(c) are delocalized even in the thermodynamic limit, because the conductance of the localized states presents exponential dependence on the system length and should be zero when $L=10^5$. This statement can be further verified in Fig.~\ref{fig:two}(d), where the averaged conductance $\langle G \rangle$ is plotted for several typical energies as a function of $L$ with other parameters identical to those in Fig.~\ref{fig:two}(c). It clearly appears that $\langle G \rangle$ fluctuates around a certain value with increasing $L$ and does not tend to zero. Therefore, we conclude that these electronic states are truly delocalized and an MIT could be driven in the two-leg ladder by increasing $t_d$.

\begin{figure}
\includegraphics[width=0.45\textwidth]{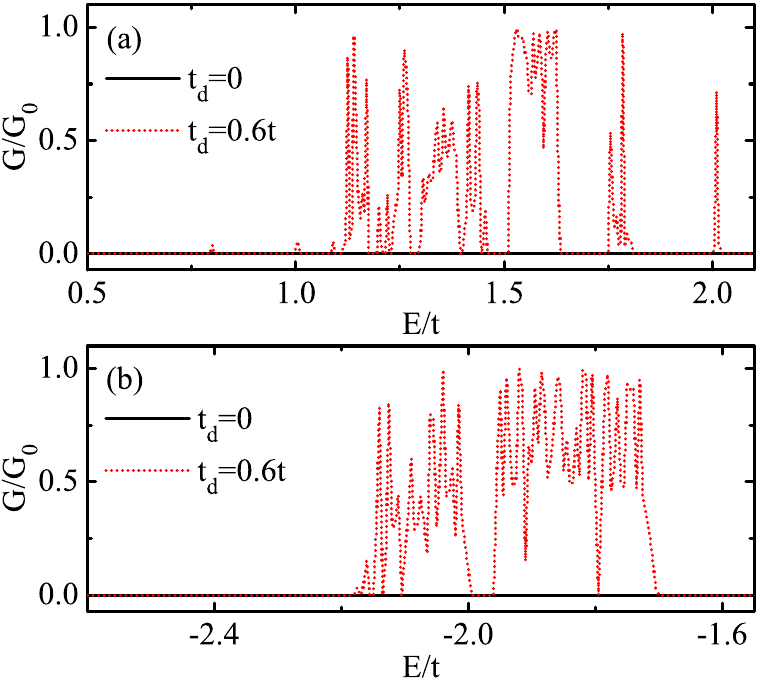}
\caption{\label{fig:three}(Color online) Energy-dependent $G$ of the two-leg ladder by coupling two distinct AA chains with $t_d=0$ (solid line) and $t_d=0.6t$ (dotted line). (a) $W_1=2.2t$, $W_2=3t$, and $\alpha= 1 /{(10 \pi)}$. (b) $W_1=t$, $W_2=4t$, and $\alpha= {(1+\sqrt 5)} /2$. Other parameters are $S=2$, $L=10^5$, and $\lambda= t$. The results are calculated from a single QP sample.}
\end{figure}

\begin{figure}
\includegraphics[width=0.45\textwidth]{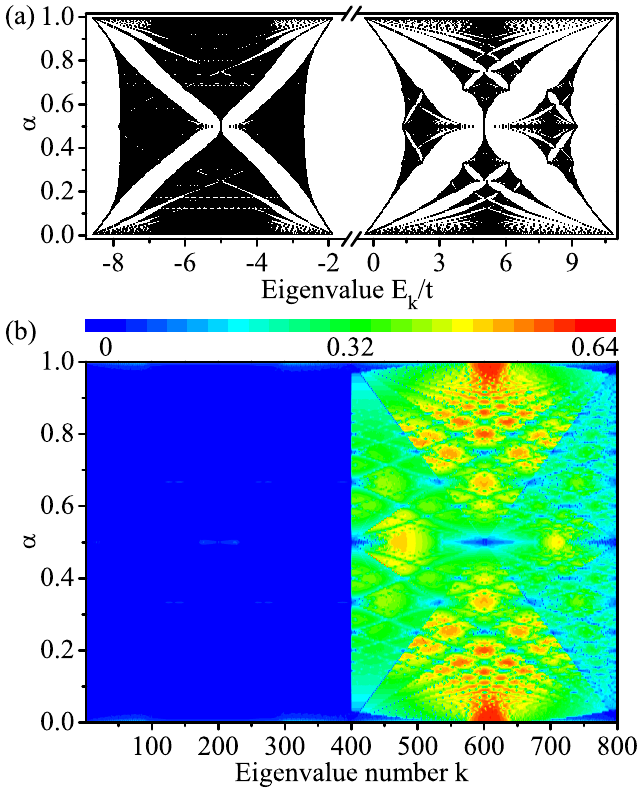}
\caption{\label{fig:four}(Color online) (a) Eigenvalue spectrum of the two-leg ladder by varying the irrational number $\alpha$ within $(0,1)$. (b) 2D plot of $P_r$ versus the eigenvalue number $k$ [see the horizontal coordinate $E_k$ in (a)] and $\alpha$. Larger $k$ refers to higher $E_k$. Here, $L=400$ and other parameters are the same as those in Fig.~\ref{fig:two}(c). All these results are performed for a single QP sample, and are similar when other QP samples are considered or by coupling two identical AA chains with $W_1=W_2$.}
\end{figure}

To further substantiate the NNN hopping-induced MIT and the stability of the extended states in the region of large $t_d$, we then investigate the localization properties of the two-leg ladder with various model parameters, as illustrated in Figs.~\ref{fig:three} and \ref{fig:four}. Firstly, we consider other values of $W_j$'s, small $\lambda$, and large $\alpha$. Figure~\ref{fig:three}(a) shows the conductance $G$ versus $E$ by coupling two localized AA chains with $W_1=2.2t$ and $W_2=3t$, while Fig.~\ref{fig:three}(b) plots $G$ versus $E$ for large irrational number $\alpha= {(1+\sqrt 5)} /2$, with the small interchain hopping integral $\lambda=t$. It can be seen that all electronic states are localized with $G=0$ in the absence of $t_d$ [see the solid lines in Figs.~\ref{fig:three}(a) and \ref{fig:three}(b)] and the extended states of nonzero conductance could emerge when $t_d$ becomes large [see the dotted lines in Figs.~\ref{fig:three}(a) and \ref{fig:three}(b)]. Since the band structure of the two-leg ladder is sensitive to both $\alpha$ and $W_j$'s, the position of the extended states is distinct from each other when $\alpha$ or $W_j$ is different.

Secondly, we consider the irrational number $\alpha$ within a wide range. Figure~\ref{fig:four}(a) plots the eigenvalue spectrum of the two-leg ladder by changing $\alpha$ from $0$ to $1$. It clearly appears that there exist two Hofstadter-like butterflies in the energy spectrum, although the system is composed of two distinct AA chains with different $W_j$'s. The left subband exhibits the compressed configuration of the Hofstadter's butterfly due to the small width $6.7t$ (the original Hofstadter's butterfly was reported at $W=2t$ with width $8t$),\cite{hdr1} and the right subband shows the expanded version with width $11.2t$.

To illustrate the localization properties at different $\alpha$, it is convenient to calculate the participation ratio $P_r$:\cite{js1}
\begin{eqnarray}
P_r= \frac {1} {SL \sum_{j=1} ^S \sum_{n=1} ^L |\psi _{j,n} |^4 }, \label{eq12}
\end{eqnarray}
which measures the extension of the wavefunction. Figure~\ref{fig:four}(b) shows the corresponding 2D plot of $P_r$ versus the eigenvalue number $k$ and $\alpha$. It can be seen that $P_r$ is very small in the left subband ($1\leq k \leq 400$) and the wavefunctions are confined within a small region. Our further calculations reveal that $P_r$ decreases with increasing $L$, indicating that all electronic states are localized in the left subband.\cite{gam1} In contrast, $P_r$ is usually very large in the right subband ($401\leq k \leq 800$) and the wavefunctions can spread over the entire system for all investigated values of $\alpha$, except for some electronic states which locate at the band edge and the corresponding $P_r$ remains quite small [see the four corners of the right subband in Fig.~\ref{fig:four}(b)]. The small $P_r$ in these corners may originate from the fact that: (1) the QP character of the system is not obvious for short system length $L=400$ when $\alpha$ is very close to an integer; (2) the electronic states in the band edge are isolated from others and cannot form a band, and thus become fragile. However, we find that the large $P_r$ is independent on $L$ (data not shown), implying that the electronic states are delocalized in the right subband. Besides, one can see that $P_r$ exhibits clustering patterns in the right subband and is extremely large within these regions, owing to the highly degenerate energy levels. Therefore, our results demonstrate that the extended states still exist in the case of large $t_d$ for all investigated values of $W_j$, $\lambda$, and $\alpha$, and thus the NNN hopping-induced MIT is general. Since the localization properties of the system does not depend upon $\lambda$, in the following we fix the interchain hopping integral as $\lambda = t$ without loss of generality.

\subsection{Delocalization induced by the dimensionality}

\begin{figure}
\includegraphics[width=0.45\textwidth]{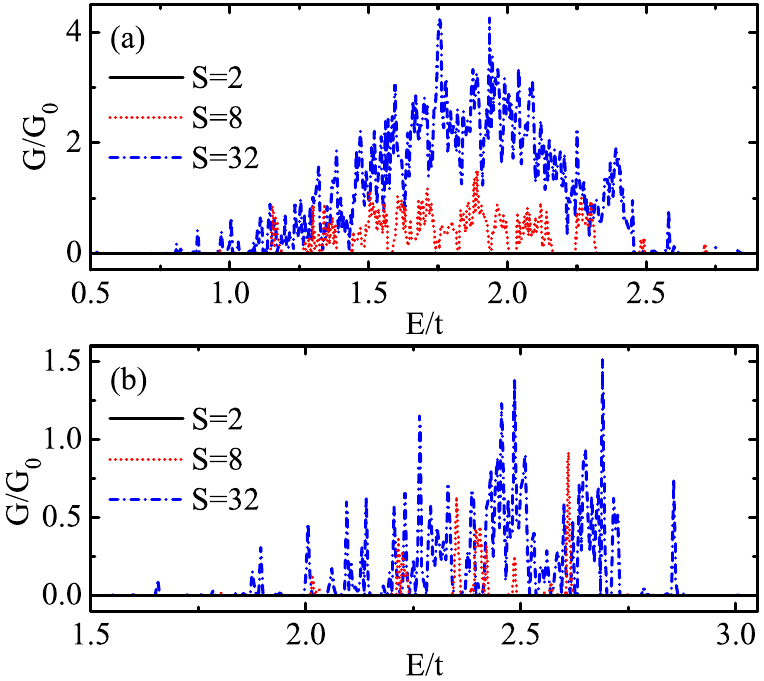}
\caption{\label{fig:five}(Color online) $G$ versus $E$ of several quasi-1D QP systems with different width $S$. (a) $W_j=2.4t$ for all $j$'s. (b) $W_j=t$ when $j$ is odd and $W_j=4t$ when $j$ is even. Other parameters are $L=10^5$, $\lambda= t$, and $t_d=0.2t$. The results are performed for a single QP sample.}
\end{figure}

Besides the NNN hopping, it can be seen from Eq.~(\ref{eq10}) that the hopping integrals of some decoupled chains can be enhanced with increasing the system width $S$ in the presence of $t_d$ and the extended states can emerge in the case of large $S$ when $W<W_c$. Figure~\ref{fig:five}(a) displays the conductance $G$ versus $E$ for several quasi-1D QP systems which are composed of identical AA chains with all $W_j=2.4t$ and $L=10^5$. One notices that $G$ is zero within the numerical accuracy for $S=2$ [see the solid line of Fig.~\ref{fig:five}(a)] and is progressively increased with increasing $S$ as expected [see the dotted and dash-dotted lines of Fig.~\ref{fig:five}(a)], implying the occurrence of MIT in the quasi-1D QP systems. The underlying physics for the extended states observed in large $S$ can be ascribed to the dimensionality effects together with the NNN hopping. Just as the two-leg ladder model, the NNN hopping integral could also generate additional propagating channels with various hopping integrals in the quasi-1D systems [Eq.~(\ref{eq10})]. These additional propagating channels will interact with the original ones at $t_d=0$ and change the transport properties of the quasi-1D system with both localized and delocalized channels. Since both the number of the delocalized channels and the conductance of the original ones increase with $S$, the system's conductance can become larger with increasing $S$. This dimensionality effects-induced MIT is a general phenomenon and can still take place in the QP systems by coupling distinct AA chains of different $W_j$'s, as illustrated in Fig.~\ref{fig:five}(b), where the conductance $G$ is shown for several quasi-1D QP systems by coupling two kinds of AA chains. Besides, the conductance profiles exhibit several gaps for small $S$ [see the dotted lines of Figs.~\ref{fig:five}(a) and \ref{fig:five}(b)] and these gaps tend to disappear for large $S$, because the energy spectra of the delocalized channels are different and can distribute over a wide energy range with increasing $S$. Therefore, we conclude that an MIT could be also driven in the quasi-1D QP systems by increasing the system width.

\subsection{Localization properties of the 2D QP system and the phase diagram}

\begin{figure}
\includegraphics[width=0.47\textwidth]{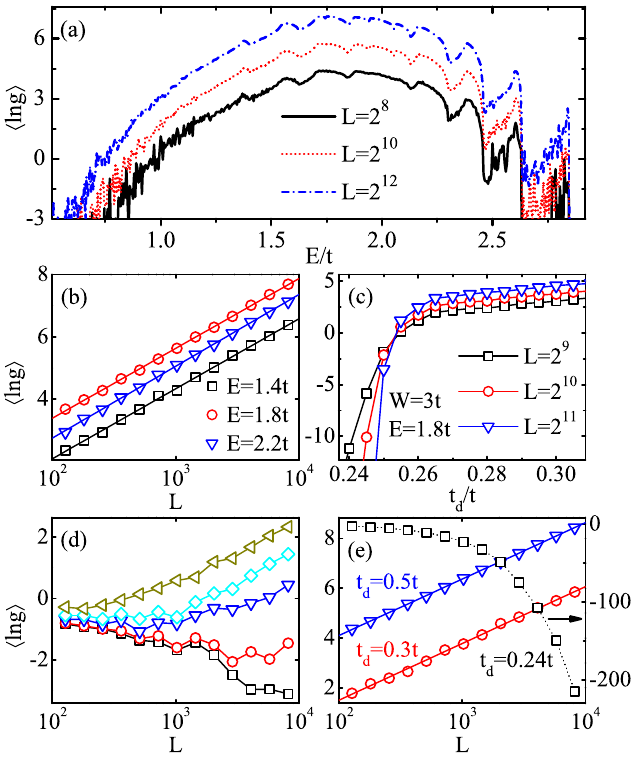}
\caption{\label{fig:six}(Color online) Scaling properties of $\langle \ln g \rangle$ for the 2D QP system with $S=L$ and all $W_j=W$. (a) $\langle \ln g \rangle$ versus $E$ for different $L$ and (b) $\langle  \ln g \rangle$ versus $L$ for several values of $E$ with $W=2.4t$ and $t_d=0.2t$. (c) $\langle \ln g \rangle$ versus $t_d$ for distinct $L$, (d) $\langle \ln g \rangle$ versus $L$ for various $t_d$ which is very close to the critical value $t_d^c$ (see text), and (e) $\langle \ln g \rangle$ versus $L$ for several $t_d$ which is distant from $t_d^c$, by fixing $W=3t$ and $E=1.8t$. In Fig.~\ref{fig:six}(d), the squares, circles, down triangles, diamonds, and left triangles denote $t_d=0.2508t$, $0.251t$, $0.252t$, $0.253t$, and $0.255t$, respectively. In Fig.~\ref{fig:six}(e), the circles and down triangles represent, respectively, $t_d=0.3t$ and $0.5t$ with their vertical axis on the left side, while the squares correspond to $t_d=0.24t$ with its vertical axis on the right side (as indicated by the arrow). The solid lines in (b) and (e) are the linear fitting curves $\langle \ln g \rangle \propto \beta_0 \ln L $ with $\beta _0$ being about $1$. The results are averaged over $10^4$ QP samples.}
\end{figure}

We then study the localization phase transition and the scaling properties of the 2D QP system with $L=S$. For the purpose of scaling, the contact resistance is subtracted from $G$ to yield the Thouless conductance $g$, which is expressed as\cite{ds1,sk2}
\begin{eqnarray}
\frac 1 g = \frac {G_0} {2G} - \frac 1 {2N}. \label{eq13}
\end{eqnarray}
Here, N is the number of propagating channels in the leads at Fermi energy $E$ with $1/(2N)$ the contact resistance. Then, the scaling function\cite{ae1,sk2}
\begin{eqnarray}
\beta=\frac {d\langle \ln g \rangle} {d \ln L} \label{eq14}
\end{eqnarray}
is numerically evaluated and is used to determine the localization and scaling properties. $\beta<0$ and $\beta>0$ correspond to the insulator and the metal, respectively.

Figure~\ref{fig:six}(a) shows the energy-dependent $\langle \ln g \rangle$ for the 2D QP system with several values of $L$. It can be seen that $\langle \ln g \rangle$ increases with $L$ in a wide energy range and exhibits strong fluctuations in the band edge, owing to the fractal energy spectrum of the AA model. The scaling behavior of $\langle \ln g \rangle$ on $L$ is plotted in Fig.~\ref{fig:six}(b) for several typical energies. It clearly appears that $\langle \ln g \rangle$ increases monotonically with $L$ and the dependence of $\langle \ln g \rangle$ on $L$ can be well fitted by a simple function $\langle \ln g \rangle \propto \beta_0 \ln L $, with the exponent $\beta_0$ being about 1. This implies that the 2D QP system presents truly metallic behavior in the thermodynamic limit.

Figure~\ref{fig:six}(c) displays $\langle \ln g \rangle$ versus $t_d$ for several $L$. One can see that $\langle \ln g \rangle$ increases with $t_d$, due to the emergence of additional delocalized channels and the enhancement of the conductance of the original ones in the 2D system. Besides, the curves $\langle \ln g \rangle$-$t_d$ for different $L$ intersect at a critical value $t_d^c$ of the NNN hopping integral. For $t_d<t_d^c$, $\langle \ln g \rangle$ decreases with $L$ and corresponds to the localized regime; whereas for $t_d>t_d^c$, $\langle \ln g \rangle$ increases with $L$ and the system is in the metallic regime. This feature can be further observed in Figs.~\ref{fig:six}(d) and \ref{fig:six}(e), where $\langle \ln g \rangle$ is shown as a function of $L$ with $t_d$ close to and distant from the critical value $t_d^c$, respectively. It can be seen that there exist fluctuations in the curve $\langle \ln g \rangle$-$L$ in the region of small conductance [see the squares, circles, and down triangles in Fig.~\ref{fig:six}(d)]. The curve $\langle \ln g \rangle$-$L$ is smooth in the region of large conductance [see the left triangles in Fig.~\ref{fig:six}(d)] and can be also fitted well by the function $\langle \ln g \rangle \propto \beta_0 \ln L $ with $\beta_0\approx1$ when $\langle \ln g \rangle$ becomes larger [see the circles and down triangles in Fig.~\ref{fig:six}(e)].

\begin{figure}
\includegraphics[width=0.45\textwidth]{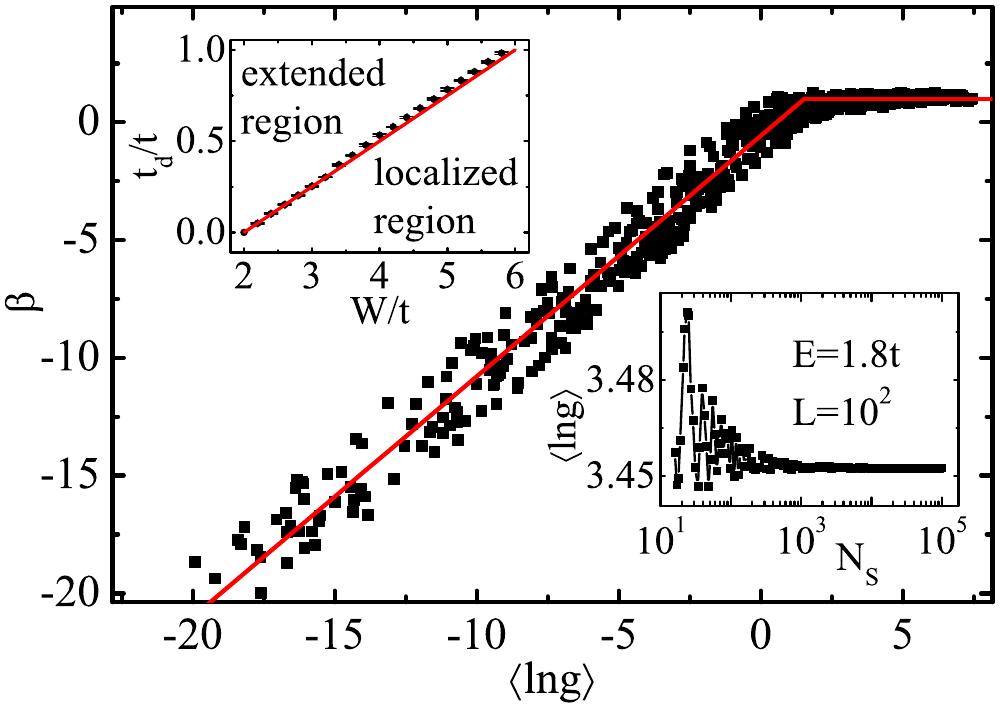}
\caption{\label{fig:seven}(Color online) Main frame: scaling function $\beta$ versus $\langle \ln g \rangle$ for the 2D QP system. $\beta$ is calculated from a wide range of $W$, $t_d$, and $E$. Here, the length is taken as $L>10^2$ to capture the QP character of the system, the step of the length is chosen as $d \ln L=0.2$, and $10^4$ QP samples are considered. The oblique line is the fitting curve of $\beta \propto 1.02 \langle \ln g \rangle$ for $\langle \ln g \rangle<0$ and the horizontal line is $\beta=1$. The top inset displays the MIT phase diagram in the ``$W$-$t_d$'' space (black symbols), which is extracted from Fig.~\ref{fig:six}(c), and the solid line denotes the curve $W=2( t+ 2 t_d)$. The bottom inset shows $\langle \ln g \rangle$ as a function of the sample number $N_S$ with $E=1.8t$, $L=10^2$, and other parameters being identical to those in Fig.~\ref{fig:six}(a).}
\end{figure}

Figure~\ref{fig:seven} plots the scaling function $\beta$ versus $\langle \ln g \rangle$ and the bottom inset shows the dependence of $\langle \ln g \rangle$ on the sample number $N_S$ for typical parameters. It clearly appears that the oscillating amplitude of $\langle \ln g \rangle$ declines quickly with increasing $N_S$ and $\langle \ln g \rangle$ saturates when $N_S > 300$ (see the bottom inset of Fig.~\ref{fig:seven}). For instance, the ratio of the difference between $\langle \ln g \rangle$ at $N_S = 294$ and its saturated value to the saturated value is 0.08\%, where the saturated value of $\langle \ln g \rangle$ is prescribed at $N_S = 10^5$. Here, $\beta$ is calculated from $10^4$ samples to avoid spurious effects and the length is taken as $L>10^2$ to capture the QP character of the system. It can be seen from Fig.~\ref{fig:seven} that all data points tend to construct a single curve of $\beta - \langle \ln g \rangle$. For small conductance, the scaling function can be linearly fitted by a simple function $\beta \propto 1.02 \langle \ln g \rangle$ (see the oblique line), in good agreement with the scaling theory of localization,\cite{ae1} implying that the conductance falls off exponentially with increasing $L$ in the localized regime. For large conductance, the scaling function is approximately $1$ (see the horizontal line of $\beta=1$), indicating that the conductance increases linearly with $L$. This point is contrary to the scaling theory of localization that the scaling function should be less than $0$ for the 2D disordered systems.\cite{ae1} The underlying physics can be attributed to the fact that the system we studied is QP but not disordered. In this situation, the delocalized channels as well as their conductances will be increased by increasing $L$.

However, there exists fluctuation in the curve $\beta - \langle \ln g \rangle$, especially in the regime of small conductance. This phenomenon is mainly due to the highly fractal energy spectrum of the QP system. One can see from Fig.~\ref{fig:six}(a) that $\langle \ln g \rangle$ fluctuates very strongly for small $\langle \ln g \rangle$, in sharp contrast to the situation that it is smooth for large $\langle \ln g \rangle$, for whatever the system size is. These fluctuations can be also identified in Fig.~\ref{fig:six}(d) as discussed above. The top inset of Fig.~\ref{fig:seven} displays the MIT phase diagram in the ``W-t$_d$'' space (see the black symbols), which is subtracted from Fig.~\ref{fig:six}(c). It can be seen that the critical NNN hopping integral $t_d$ to observe metallic states increases monotonically with $W$ and the curve of $t_d$-$W$ can be fitted by $W=2(t+2t_d)$.

\subsection{Statistical properties of the 2D system}

\begin{figure}
\includegraphics[width=0.37\textwidth]{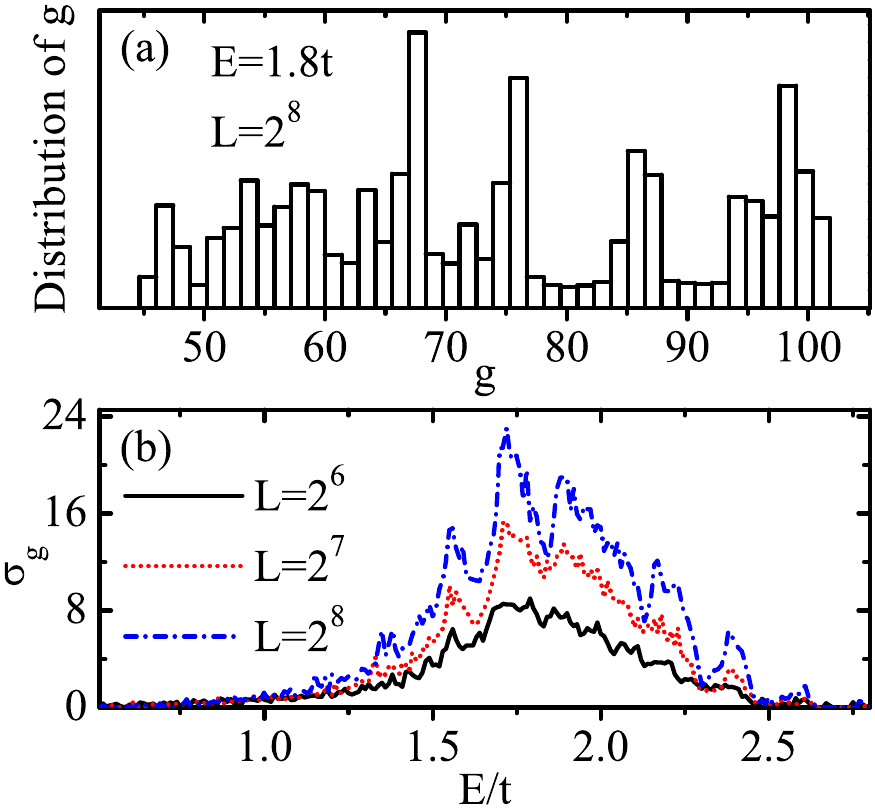}
\caption{\label{fig:eight}(Color online) Statistical properties of the Thouless conductance for the 2D QP system. (a) Distribution of $g$ and (b) energy-dependent standard deviation $\sigma_g =\sqrt{\langle g^2 \rangle - \langle g \rangle ^2}$. The results are obtained from $10^5$ QP samples and other parameters are identical to those in Fig.~\ref{fig:six}(a). }
\end{figure}

Finally, we investigate the statistical properties of the Thouless conductance of the 2D QP system. Figures~\ref{fig:eight}(a) and \ref{fig:eight}(b) show, respectively, the distribution of $g$ for typical parameters and the energy-dependent standard deviation $\sigma_g =\sqrt{\langle g^2 \rangle - \langle g \rangle ^2}$ for three system sizes with $N_S=10^5$. It can be seen from Fig.~\ref{fig:eight}(a) that the distribution of $g$ is extremely complicated and is not normal, which is different from previous studies in disordered systems.\cite{sk1,sk3} Because of the cosine modulation of the on-site energies and the quasiperiodicity, we find that the conductance presents oscillating behavior with various amplitudes when the sample is successively chosen from an infinite AA chain by using the sliding window strategy (data not shown). Besides, we can see that the probability is quite large or small when the conductance is around specific values. As a result, the distribution of $g$ will deviate from the normal one.

By inspecting Fig.~\ref{fig:eight}(b), one notices the following features: (1) the magnitude of $\sigma_g$ is comparable to $\langle g \rangle$ [see Fig.~\ref{fig:six}(a) and $\exp(\langle \ln g \rangle)$ can be viewed as approximation of $\langle g \rangle$], where $\sigma_g$ is large around the band center and becomes very small within the band edge. It can be also observed that $\sigma_g$ becomes greater when $\langle g \rangle$ is larger, indicating that the fluctuation of the conductance will be stronger for larger conductance. (2) $\sigma_g$ is enhanced by increasing $L$ around the band center, where the conductance increases with $L$ [see Fig.~\ref{fig:six}(a)] and the system presents metallic behavior. This is different from the universal conductance fluctuation theory that $\sigma_g$ should be independent of the system size. (3) There are several dips in the profile of $\sigma_g - E$, similar to that observed in Fig.~\ref{fig:six}(a).

\section{Conclusions\label{sec4}}

In summary, we investigate the localization phase transition and the scaling properties of both quasi-one-dimensional and two-dimensional quasiperiodic systems, which are constructed by coupling several Aubry-Andr\'{e} chains along the transverse direction. By employing the Landauer-B\"{u}ttiker formula and recursive Green's function method, the two-terminal conductance is calculated for the quasiperiodic systems with next-nearest-neighbor hopping. The numerical results indicate that a metal-insulator transition could be produced in these quasiperiodic systems in parameter space by adjusting the next-nearest-neighbor hopping integral and the number of chains. These results are general and hold by coupling distinct AA chains with different strengths of the on-site potentials and other model parameters. We find that the energy spectrum of the two-leg ladder model exhibits two Hofstadter-like butterflies. Besides, we show from the finite-size scaling that the transport properties of the two-dimensional quasiperiodic system can be described by the single Thouless conductance and the scaling function can reach the value 1, contrary to the scaling theory of localization.

\section*{Acknowledgments}

We thank Yan-Yang Zhang for useful discussions. This work was
financially supported by NBRP of China (2012CB921303 and
2012CB821402), NSF-China under Grant Nos. 11274364 and 91221302, and
PDSF-China under Grant No. 2013M540153.

\end{document}